\documentclass[aps,prc,amsmath,nofootinbib,twocolumn,showpacs]{revtex4}
\usepackage{graphicx}
\begin{document}
\title{Hyperonic mixing in five-baryon double-strangeness hypernuclei
in a two-channel treatment }
\author{D.\ E.\ Lanskoy}
\email{lanskoy@sinp.msu.ru} \affiliation{Institute of Nuclear
Physics, Moscow State University, 119992 Moscow, Russia}
\author{Y.\ Yamamoto}
\email{yamamoto@tsuru.ac.jp} \affiliation{Physics Section, Tsuru
University, Tsuru, Yamanashi 402-8555, Japan}
\date{\today}

\begin{abstract}
Properties of hypernuclei $_{\Lambda \Lambda }^5$H and $_{\Lambda
\Lambda }^5$He are studied in a two-channel approach with explicit
treatment of coupling of channels $^3\text{Z}+\Lambda+\Lambda$ and
$\alpha+\Xi$. Diagonal $\Lambda\Lambda$ and coupling
$\Lambda\Lambda-\Xi N$ interactions are derived within $G$-matrix
procedure from Nijmegen meson-exchange models. Bond energy $\Delta
B_{\Lambda\Lambda}$ in $_{\Lambda \Lambda }^5$He exceeds
significantly that in $_{\Lambda \Lambda }^5$H due to the channel
coupling. Diagonal $\Xi\alpha$ attraction amplifies the effect,
which is sensitive also to $\Lambda$-core interaction. The
difference of the $\Delta B_{\Lambda\Lambda}$ values can be an
unambiguous signature of the $\Lambda\Lambda-\Xi N$ coupling in
$\Lambda\Lambda$ hypernuclei. However, improved knowledge of the
hyperon-nucleus potentials is needed for quantitative extraction
of the coupling strength from future data on the $\Lambda\Lambda$
hypernuclear binding energies.
\end{abstract}
\maketitle

\section{\label{s1}Introduction}

Mixing of different baryonic states in nuclear systems remains a
topic rather exciting but being far from sophisticated
understanding. A long time ago, it was speculated that a nucleus
is not generally a pure nucleonic system. Due to
$NN\leftrightarrow N\Delta $ and $NN\leftrightarrow \Delta \Delta
$ transitions, a nuclear wave function incorporates some admixture
of states with one (or even more) $\Delta $ baryon(s)
\cite{Green}. Strictly speaking, also heavier baryons should be
taken into account.

Admixtures of $\Sigma $ states in $\Lambda $ hypernuclei probably
are more important for hypernuclear dynamics. The $\Sigma \Lambda
$ mass difference (80 MeV) is rather less than the $\Delta N$
difference (300 MeV). Moreover, pionic exchange between $\Lambda $
and $N$ necessarily gives rise to virtual $\Sigma $ because of
zero $\Lambda $ isospin. It was suggested a long ago \cite{Bod}
that the $\Lambda N-\Sigma N$ coupling is responsible for the
so-called $_\Lambda ^5$He overbinding problem, which was confirmed
recently in a consistent study of $s$-shell hypernuclei
\cite{Ak00,NAS}. Probably, the $\Lambda N-\Sigma N$ coupling plays
the crucial role in binding of hypertriton \cite{MG}. The coupling
contributes greatly to the $\Lambda $ binding in nuclear matter
\cite{Yam94}.

However, there are no direct unambiguous measurements of the
baryonic admixtures so far. Various processes are considered to
detect $\Delta $ admixtures in ordinary nuclei, but discrimination
between $\Delta $ components preexisting in a nucleus and $\Delta
$ baryons produced in a reaction is a difficult problem (for some
recent attempts, see \cite {Pas,Bys}). For contributions of
$\Lambda N-\Sigma N$ mixing to $\Lambda $ hypernuclear binding
energies, alternative dynamical pictures without explicit $\Sigma
$ admixtures usually exist. For instance, when a single channel
$\Lambda N$ interaction is described in terms of correlated
two-pion exchange, it is probable that the dynamics of virtual
intermediate $\Sigma $'s is included implicitly.
Also effective $\Lambda NN$ force can mimic effects of the $%
\Lambda N-\Sigma N$ coupling without explicit $\Sigma $ degrees of freedom.

Several other implications of the $\Lambda N-\Sigma N$ coupling providing in
principle ways to identify the $\Sigma $ admixtures, have been discussed. It
was suggested that measurement of hypernuclear magnetic moments is promising
for this aim \cite{FDG}. The probability of the rare $\pi ^{+}$ channel of
hypernuclear weak decay is sensitive to $\Sigma ^{+}$ admixture \cite{vH,GT}.
The coupling can lead in some cases to the reverse order of spin doublet
levels \cite{G94}. Also production of neutron-rich $\Lambda $ hypernuclei
from $(K^{-},\pi ^{+})$ and $(\pi ^{-},K^{+})$ reactions can proceed via $%
\Sigma^-$ admixture as a doorway state \cite{TL01}. Relevant data
are not available so far. Moreover, in all of these cases, some
background effects [as meson charge exchange in the $\pi ^{+}$
decay as well as in the $(K^{-},\pi ^{+})$
and $(\pi ^{-},K^{+})$ reactions] occur, which can hinder detection of $%
\Sigma $ admixture.

The $\Lambda \Lambda -\Xi N$ coupling in double-strangeness
hypernuclei is of particular interest, since the relevant mass
difference (about 25 MeV) is lowest among all possible known
baryonic couplings. However, experimental information on
double-strangeness systems is rather scarce so far, and no
definite knowledge of the coupling exists yet.

Theoretically, the $\Lambda \Lambda -\Xi N$ coupling in $\Lambda \Lambda $
hypernuclei\footnote{Strictly
speaking, the term ``$\Lambda \Lambda $ hypernucleus''
in this context means a state of an $S=-2$ hypernucleus with
dominant $\Lambda \Lambda $ component.} have been considered by several
groups \cite{G94,Ak94,Yam94,DGM,G97,YaN,G03,Ak03}. Mostly, hypernuclei
observed experimentally have been studied, namely, $_{\Lambda \Lambda }^6$He
\cite{G94,G97,YaN,G03}, $_{\Lambda \Lambda }^{10}$Be, $_{\Lambda \Lambda
}^{13}$B \cite{Yam94,YaN}. With meson-exchange coupling potentials,
probabilities of $\Xi $ admixtures less than 1\% were obtained.
Contributions of the coupling to the binding energies are as small as
several tenths of MeV except the case of extremely strong $\Lambda \Lambda $
attraction providing $\Lambda \Lambda $ bound state \cite{G97,G03}, when the
contribution
can reach several MeV. Much larger coupling has been obtained \cite{YaN}
within a quark model predicting free bound $H$ dibaryon. In this case, not
only $\Xi N$, but also $\Sigma \Sigma $ component is of a great weight (more
than 10\%).

Myint and Akaishi \cite{Ak94} argued that the $\Lambda \Lambda
-\Xi N$ coupling is considerably enhanced in five-baryon
hypernucleus $_{\Lambda \Lambda }^5$H. A proton, appearing from
the $\Lambda \Lambda \rightarrow \Xi ^{-}p$ transition, can be
bound rather strongly in the $\alpha $ particle. Thus, the
difference between the thresholds of channels $^3\text{H}+\Lambda
+\Lambda $ and $^4\text{He}+\Xi ^{-}$ is reduced to 8 MeV from 29
MeV for free $\Lambda \Lambda $ and $\Xi ^{-}p$ pairs. Myint and
Akaishi \cite{Ak94} obtained 1\% for the $\Xi^-$ admixture
probability and 0.5 MeV for the binding excess appearing due to
the coupling. These values are larger than those typically
obtained by other authors for other hypernuclei, but still small
to provide more or less unambiguous signature of the coupling.

In the studies performed in the 1990s, Nijmegen hard-core model D
(NHC-D) \cite{ND} has been used popularly as a standard
meson-theoretical model for $S=-2$ interactions. The reason was
that this model is compatible with strong $\Lambda \Lambda $
attraction ($\Delta B_{\Lambda \Lambda }=4$--5 MeV) supported by
earlier data on $\Lambda \Lambda$ hypernuclei \cite{Dal,Ao}. This
strong $\Lambda \Lambda $ attraction of NHC-D is due to its
specific feature that only the scalar singlet is taken into
account. In the cases of the other Nijmegen models incorporating
the whole scalar nonet, the meson-exchange parts of the $\Lambda
\Lambda$ interactions are much weaker than those of NHC-D. In the
case of the hard-core model F (NHC-F) \cite{NF}, for instance, the
strength of that part is about a half of NHC-D. Of course, in
these models the hard-core radii can be treated as adjustable
parameters to reproduce any strength of $\Lambda \Lambda$
interactions. Then, it is difficult to discriminate between strong
and weak meson-exchange attraction compensated by small and large
hard-core radii, respectively, in $\Lambda \Lambda$ single-channel
treatment.

On the other hand, the $\Lambda \Lambda -\Xi N$ coupling
of NHC-D is relatively weak because of small contributions of strange mesons,
which leads to small $\Xi $ admixtures appeared in structure calculations of
double-$\Lambda$ hypernuclei \cite{Yam94}. In the case of NHC-F, for instance,
the $\Lambda \Lambda -\Xi N$ coupling has been known to be stronger by about
two times than that of NHC-D.
As shown later, the strength of the $\Lambda \Lambda -\Xi N$ coupling of
NHC-D is the weakest among the various Nijmegen models due to
the above-mentioned specific feature.

Since the discovery of Nagara event identified uniquely as $^{\
6}_{\Lambda \Lambda}$He \cite{Nag} in 2001, $\Lambda \Lambda $
attraction is believed to be rather less ($\Delta B_{\Lambda
\Lambda } \approx 1 $ MeV). On the basis of this new datum, Hiyama
\textit{et al.}\ have performed the systematical analysis for
light double-$\Lambda$ hypernuclei \cite{Hiya02}. The strength of
obtained $\Lambda \Lambda$ interaction is very similar to the
meson-exchange part of NHC-F. Even if NHC-D is used, it is
possible to reproduce $\Delta B_{\Lambda \Lambda }(^{\ 6}_{\Lambda
\Lambda}\text{He}) \approx 1 $ MeV by taking a larger value of its
hard-core radius appropriately. However, weak $\Lambda \Lambda$
attraction consistent with $\Delta B_{\Lambda \Lambda }\approx 1 $
MeV can be obtained more plausibly by the other Nijmegen models.
Therefore, it is likely that their stronger $\Lambda \Lambda -\Xi
N$ coupling interactions are more realistic and the mixing in
$\Lambda \Lambda $ hypernuclei is more significant than it was
speculated earlier from NHC-D.

Very recently, Myint \textit{et al.}\ \cite{Ak03} performed a new
study of the five-baryon $\Lambda \Lambda $ hypernuclei
($_{\Lambda \Lambda }^5$H as well
as $_{\Lambda \Lambda }^5$He) and showed that the coupling effect in $%
_{\Lambda \Lambda }^5$He exceeds considerably that in $_{\Lambda \Lambda }^5$%
H. Since $\Xi ^0$ is lighter than $\Xi ^{-}$ by 6.4 MeV, the
threshold difference of channels $^3\text{He}+\Lambda +\Lambda $
and $^4\text{He}+\Xi ^0$ is as small as 2.4 MeV. So the $\Xi $
admixture is probably greatest in $_{\Lambda \Lambda }^5$He, and
remarkable difference between $_{\Lambda \Lambda }^5$He and $%
_{\Lambda \Lambda }^5$H binding energies appears. Similar
suggestions have been made also in \cite {L03}.

In this paper, we study properties of $_{\Lambda \Lambda }^5$H and $%
_{\Lambda \Lambda }^5$He with diagonal $\Lambda \Lambda $ and coupling $%
\Lambda \Lambda -\Xi N$ $G$-matrix interactions deduced from
various meson-exchange models. While Myint \textit{et al.}\
\cite{Ak03} employed a single-channel approach, in which the
coupling is involved effectively in the diagonal potential, we
solve the relevant two-channel problem explicitly. Particularly,
we incorporate the diagonal $\Xi \alpha $ potential into the
calculation and emphasize that effect of this potential is rather
important. We investigate also the role of the other inputs
($\Lambda -{}^3$Z and $\Lambda \Lambda $ potentials).

In Sec.\ \ref{s2} we describe our model, choice of
phenomenological hyperon-core potentials, and also $G$-matrix
derivation of the diagonal $\Lambda\Lambda$ and coupling
$\Lambda\Lambda-\Xi N$ effective interaction. Section \ref{s3}
contains presentation and discussion of our results. Also
comparison of our approach with that of Myint \textit{et al.}\
\cite{Ak03} is outlined. Section \ref{s4} is devoted to concluding
remarks and some outlook.

\section{\label{s2}Model and interactions}
\subsection{\label{s2a}Model}
Accurate five-body calculations for hypernuclei $_{\Lambda \Lambda
}^5$H and $_{\Lambda \Lambda }^5$He are available so far only in
the single-channel approach without baryonic mixings \cite{Nem}.
In \cite{Ak94,Ak03}, three-body models were utilized with free
two-baryon $S=-2$ potentials (more precisely, their
phase-equivalents). Since we also do not solve the five-body
problem directly, we use effective ($G$-matrix) two-body
$\Lambda \Lambda -\Lambda\Lambda$ and $%
\Lambda \Lambda -\Xi N$ interactions instead of free-space ones.
Here, we adopt the Hartree-Fock (HF) description, which is not
only simpler, but also is suitable for the $G$-matrix formalism
defined in a single-particle model space.

Because density distributions of the lightest nuclei are known to
be described poorly in the HF approximation, we do not apply the
HF description for all the five baryons. Instead, we treat the
hypernuclei as $^3\text{Z}+\Lambda +\Lambda $ and $^4\text{He}+\Xi
$ in the $\Lambda \Lambda $ and $\Xi $ channels, respectively. The
nuclear cores are treated as inert, and hyperon-nucleus
interactions are described fully by $\Lambda -{}^3$Z and $\Xi
-{}^4$He potentials. This HF approach is similar to that in
\cite{Car}, but it is extended to incorporate the second channel.

We adopt the wave function of the five-baryon hypernucleus as
$\Psi=\psi_3\psi_\Lambda(\mathbf{r}_1)\psi_\Lambda(\mathbf{r}_2)+
\psi_3\psi_N(\mathbf{r}_1)\psi_\Xi(\mathbf{r}_2)$, where $\psi_3$,
$\psi_\Lambda$, $\psi_N$, and $\psi_\Xi$ are the three-nucleon
core wave function and single-particle wave functions of
$\Lambda$, nucleon ($N$), and $\Xi$, respectively. The $N$ and
$\Xi$ components appear from the $\Lambda\Lambda\to\Xi N$
transition. For brevity, spin parts are not shown explicitly in
the wave function above. We restrict ourselves to the single
$\psi_N$ state corresponding to the ground state $\alpha$ particle
and also to the $1s$ states of the hyperons. Since all the baryons
are in $s$ states, the kinetic energy term containing scalar
product of gradients acting on $\mathbf{r}_1$ and $\mathbf{r}_2$
\cite{Car} vanishes identically. Excluding c.m.\ motion in the
usual way and neglecting at that instance the difference between
c.m.\ positions in the two channels, we come to equations for
single-particle wave functions.

For the radial wave functions, the equations to be solved are as follows:
\begin{eqnarray}
-\frac 1{2\mu _\Lambda }\varphi ^{^{\prime \prime }}+U_\Lambda
\varphi &=&e\varphi -\sigma \chi,\label{e1}\\* -\frac 1{2\mu _\Xi
}\chi ^{^{\prime \prime }}+U_\Xi \chi &=&(e-\Delta e)\chi -\sigma
\varphi.\label{e2}
\end{eqnarray}
Here $\varphi $ and $\chi $ are the radial wave functions of $\Lambda $ and $%
\Xi $, respectively. The threshold-energy difference $\Delta e$
is taken as 8.4 (2.4) MeV for $_{\Lambda \Lambda }^5$H (%
$_{\Lambda \Lambda }^5$He). Then, $U_\Lambda =U_{c\Lambda
}+\left\langle \varphi \left| V_{\Lambda \Lambda }\right| \varphi
\right\rangle $ and $U_\Xi $ are the diagonal potentials, whereas
$\sigma \propto \left\langle \varphi \left| V_{\Lambda \Lambda
,\Xi N}\right| \varphi_N \right\rangle $ is the nondiagonal one,
where the nucleon radial wave function $\varphi_N$ is calculated
in the same potential as in \cite{Ak94}. Potentials $U_{c\Lambda
}$, $U_\Xi $, $V_{\Lambda \Lambda }$ and $V_{\Lambda \Lambda ,\Xi
N}$ correspond to $\Lambda$-core, $\Xi\alpha$, $\Lambda\Lambda$
diagonal and  $\Lambda \Lambda-\Xi N$ coupling interactions,
respectively. If $\psi_3$ and $\psi_N$ are normalized to unity,
the normalizing condition for $\varphi$ and $\chi$ is $\left(
\int_0^\infty\varphi^2dr\right)^2+\int_0^\infty\chi^2dr=1$. Since
$U_\Lambda $ and $\sigma $ depend on $\varphi $, the system must
be solved self-consistently as usually in HF approaches.

In (\ref{e2}), only one state ($\alpha +\Xi $) is kept. Generally,
it is an important problem how to take account of other states
beyond the $1s$ shell.
It was shown that the $%
\Lambda \Lambda -\Xi N$ coupling contributes to the $_{\Lambda
\Lambda }^6$He binding energy sizably (though not so much for
realistic interactions) \cite{G97,YaN,G03}, despite that all the
nucleonic $1s$ states are occupied. In our approach,
highly-excited components are renormalized into the effective
$\Lambda \Lambda$ interactions in some approximate way, as
commented in the end of Subsec.\ \ref{s2c}.

\subsection{\label{s2b}$\Lambda$- and $\Xi$-core interactions}

Our main purpose in this work is to calculate
the quantity $\Delta B_{\Lambda \Lambda }$ defined as
\begin{equation}
\Delta B_{\Lambda \Lambda }=B_{\Lambda \Lambda }-2B_\Lambda
,\label{e3}
\end{equation}
where $B_\Lambda $ and
$B_{\Lambda \Lambda }$ denote binding energies of $\Lambda$
and a pair of $\Lambda$'s in a nuclear core, respectively.
In order to obtain $\Delta B_{\Lambda \Lambda}$ values of
$^{\ 5}_{\Lambda \Lambda }$H, $^{\ 5}_{\Lambda \Lambda }$He, and
$^{\ 6}_{\Lambda \Lambda }$He, we need
$\Lambda-^{3}$H, $\Lambda-^{3}$He, and $\Lambda-^{4}$He potentials.

For $\Lambda$-core interactions, we use phenomenological prescriptions.
For $\Lambda -{}^3$Z potential $U_{c\Lambda }$ we use the Isle-type form
\cite{Kur}
\begin{equation}
U_{c\Lambda}=U_0^{\text{Isle}}(1.106\exp (-r^2/r_1^2)-\exp
(-r^2/r_2^2))\label{e4}
\end{equation}
with $r_1=1.25$ fm, $r_2=1.41$ fm.
We fit the strength to binding energies of $_\Lambda ^4$H and $%
_\Lambda ^4$He in the ground and the first excited states, and then average
the potential over the singlet and triplet states.
The obtained values are $U_0^{\text{Isle}}=322.8$ MeV ($%
_\Lambda ^4$H) and $U_0^{\text{Isle}}=338.3$ MeV ($_\Lambda
^4$He). Quantity $B_\Lambda$ (1.25 MeV in $_\Lambda ^4$H and 1.53
MeV in $_\Lambda ^4$He) in (\ref{e3}) is calculated in this
averaged potential. For the $\Lambda -{}^4$He potential, fitting
is performed to $B_\Lambda (_\Lambda ^5$He), which gives
$U_0^{\text{Isle}}=394.9$ MeV.

To address sensitivity of the results to the shape of the hyperon-nucleus
potentials, we examine also another force. Namely, $U_{c\Lambda }$ is taken
as one-range Gaussian (ORG):%
\begin{equation}
U_{c\Lambda} =U_0^{\text{ORG}}\exp (-r^2/r_0^2)\label{e5}
\end{equation}
with $r_0=1.5656$ fm and $U_0^{\text{ORG}}=-43.93$ MeV for
$_\Lambda ^5$He \cite{Das}. The fitting procedure for the
potential strengths is the same as described above. We obtain
$U_0^{\text{ORG}}=-38.38$ $(-39.78)$ MeV for $_\Lambda ^4$H
($_\Lambda ^4$He).

It is well known that the effect of the  $\Lambda N-\Sigma N$
coupling is especially important just in the $_\Lambda ^4$H and
$_\Lambda ^4$He hypernuclei. Thus, this effect may be expected to
be important also in the five-body $\Lambda\Lambda$ hypernuclei.
In our model, the $\Lambda N-\Sigma N$ coupling is incorporated
effectively at the mean-field level in the $\Lambda$-nucleus
potentials, since they are determined so as to reproduce the
experimental $\Lambda$ binding energies in $_\Lambda ^4$H,
$_\Lambda ^4$He, and $_\Lambda ^5$He. It should be noted that some
direct interplay between two types of baryonic mixing is also
possible, which is beyond the mean-field treatment. For instance,
a $\Sigma$ hyperon appeared from the $\Lambda N-\Sigma N$ coupling
can participate in further couplings (like $\Lambda\Sigma -\Xi
N$). However, since the $\Sigma$ admixtures are not larger than
2\% \cite{Ak00,NAS,HKMYY}, we omit such effects.

In our two-channel model, important roles are played by $\Xi$-core
interactions. Particularly, we study effects of the $\Xi \alpha $
potential $U_\Xi$ for $\Xi$ admixtures in $_{\Lambda\Lambda}^5$H
and $_{\Lambda\Lambda}^5$He. Unfortunately, too little is known
about $\Xi $-nucleus interaction. The $\Xi$-core attractions
comparable to $\Lambda$-core ones were deduced by analyzing the
compilation of the emulsion events of ``$\Xi$ hypernuclei'' in the
past \cite{DG}, although none of individual events was identified
unambiguously. The recent data on $^{12}$C($K^{-},K^{+})$ reaction
\cite{Fuk,Kh} give some evidence that the $\Xi $-nucleus
interaction, being attractive, is weaker by about half than the
corresponding $\Lambda $-nucleus interaction. We consider that
this information is most reliable in the present stage. We adopt
here a simple folding model to obtain the strength of $\Xi N$
interaction compatible with $\Xi $ well depth \cite{Fuk,Kh} of 14
MeV in $^{11}$B. The $\Xi \alpha$ potential is deduced by folding
of the obtained $\Xi N$ interaction with the density of $\alpha$,
and then $\Xi$ binding energies in the $\Xi \alpha $ systems are
calculated. We find no bound state of $_{\Xi ^0}^5$He, while $\Xi
^{-}$ hyperon is bound by $\alpha $ particle with $B_{\Xi
^{-}}(_{\Xi ^{-}}^5$H)$=0.4$--0.5 MeV (depending on details of the
folding procedure). This value is much greater than the
corresponding Coulomb energy (about 0.1 MeV). The rms radius
(about 6 fm) also confirms that it is a nuclear (or, maybe,
``hybrid Coulomb-assisted'' \cite{hCa}) state. We consider also
the $\Xi $ well depth in $^{11}$B of 24 MeV according to the
earlier analysis \cite{DG}. In this case, bound $_{\Xi ^0}^5$He
appears with $B_{\Xi ^0}=0.9$--1.1 MeV. The much stronger $\Xi
\alpha $ potential suggested by Filikhin and Gal \cite{FG} gives
$B_{\Xi ^0}=2.1$ MeV and thus it is probably incompatible with the
data \cite{Fuk,Kh}.

For radial dependence of $\Xi \alpha $ potential $U_\Xi $, we
adopt the Isle-type form (4) as well. We fit strength
$U_0^{\text{Isle}}$ to $B_{\Xi ^{-}}(_{\Xi^{-}}^5$H)$=0.5$ MeV
(potential Xa1). For comparison, we examine the Filikhin and Gal
\cite{FG} choice ($B_{\Xi ^0}(_{\Xi ^0}^5$He)$=2.1$ MeV, potential
Xa2). At last, we test also zero $\Xi \alpha $ potential Xa0 (to
be consistent with calculations \cite{Ak03}, we switched off also
Coulomb $\Xi ^{-}\alpha $ interaction in the last case). For
consistency with the $\Lambda$-core potentials, we prepare also
the $\Xi \alpha $ potentials in the ORG\ form with $r_0=2.145$ fm,
which corresponds to folding of a $\Xi N$ ORG-type potential with
Gaussian $\alpha $ density with reasonable range parameters.
Potential Xa3 is fitted to the same value $B_{\Xi
^{-}}(_{\Xi^{-}}^5$H)$=0.5$ MeV as potential Xa1. The strongest
ORG-type $\Xi \alpha $ potential Xa4 is fitted to
$B_{\Xi ^0}(_{\Xi ^0}^5$H)$%
=1.06 $ MeV, which is compatible with $\Xi $ well depth of 24 MeV
\cite{DG} in $^{11}$B, being rather less than the potential by
Filikhin and Gal \cite {FG} predicts. The $\Xi \alpha $ potentials
used are presented in Table \ref{t1}.

\begin{table*}[tbh]
\caption{\label{t1}Parameters of the Isle-type
($U_0^{\text{Isle}}$) and ORG ($U_0^{\text{ORG}}$) $\Xi\alpha$
potentials. Corresponding $\Xi$ binding energies are also shown.
All the quantities are in MeV.}
\begin{ruledtabular}
\begin{tabular}{cccccccc}
\multicolumn{4}{c}{Isle, Eq.\ (\ref{e4})}& \multicolumn{4}{c}{ORG,
Eq.\ (\ref{e5}), $r_0=2.145$ fm}\\
No.&$U_0^{\text{Isle}}$&$B_{\Xi^0}(^5_{\Xi^0}$He)&$B_{\Xi^-}(^5_{\Xi^-}$H)&
No.&$U_0^{\text{ORG}}$&$B_{\Xi^0}(^5_{\Xi^0}$He)&$B_{\Xi^-}(^5_{\Xi^-}$H)\\
\hline Xa1 & 158.8 & unbound & 0.50 & Xa3 & $-1.816$ &
unbound&0.50\\ Xa2 & 319.9 & 2.09 & 3.25 & Xa4 & $-3.108$ &1.06
&2.14\\ Xa0 & 0     & $-$  & $-$  & Xa0 & 0      & $-$ &$-$\\
\end{tabular}
\end{ruledtabular}
\end{table*}

\subsection{\label{s2c}Diagonal $\Lambda \Lambda$ and coupling
$\Lambda \Lambda-\Xi N$ interactions}

Let us derive the effective $\Lambda \Lambda -\Lambda \Lambda $ and $%
\Lambda \Lambda -\Xi N$ interactions suitable for our HF model
space, starting from the underlying free-space interactions. We
adopt here the $\Lambda \Lambda -\Lambda \Lambda $ and $\Lambda
\Lambda -\Xi N$ sectors of the SU(3)-invariant OBE models by
Nijmegen group; not only the hard-core models NHC-D and -F, but
also the soft-core models NSC89 \cite{NSC89} and NSC97
\cite{NSC97}. There are six versions (a--f) of the NSC97 model. In
this work, we choose the e and f versions as typical examples.

The $G$-matrix theory is most convenient to derive an effective
interaction in some model space: Our effective interactions in the
HF single-particle space are given by the $G$-matrix interactions
$G_{\Lambda \Lambda,\Lambda \Lambda}$, into which high-momentum
transfer components beyond our model space are renormalized. On
the other hand, the effective $\Lambda \Lambda -\Xi N$ coupling
interaction $G_{\Lambda \Lambda, \Xi N}$ controls the mixing of
$\Xi $ components with $\Lambda \Lambda $ states within our HF
model space.

We note here the basic feature of our model for $_{\Lambda \Lambda }^6$He
and $_{\Lambda \Lambda }^5$H ($_{\Lambda \Lambda }^5$He):
In the case of $_{\Lambda \Lambda }^6$He, where the mixing of $\Xi N$
component is out of our model space due to the Pauli effect,
the $\Lambda \Lambda$ state is described by
$G_{\Lambda \Lambda,\Lambda \Lambda}$
in the single-channel treatment.
In the case of $_{\Lambda \Lambda }^5$H ($_{\Lambda \Lambda }^5$H),
for which the $\Xi N$ mixing is treated explicitly within our model space,
the $\Lambda \Lambda$ and $\Xi N$ mixed states are described by
$G_{\Lambda \Lambda,\Lambda \Lambda}$ and $G_{\Lambda \Lambda, \Xi N}$
in the two-channel treatment.

In this work we construct simply the $G$-matrix interactions in
nuclear matter \cite{Yam94}, which can be considered approximately
as the effective interactions in our finite model space. This
nuclear-matter approach is sufficient for our purpose to explore
$\Lambda \Lambda -\Xi N$ coupling effects, considering
uncertainties of underlying free-space interactions. The
coupled-channel $G$-matrix equation for $\Lambda \Lambda $ and
$\Xi N$ pairs in symmetric nuclear matter with a Fermi momentum
$k_F$ is written as
\begin{equation}
\label{eq:Gmat}G_{cc_0}=v_{cc_0}+\sum_{c^{\prime }}v_{cc^{\prime }}{\frac{%
Q_{y^{\prime }}}{e_{yy^{\prime }}}}G_{c^{\prime }c_0}\ ,
\end{equation}
where $c$ denotes a relative state for a pair $y=(\Lambda \Lambda
)$ or $(\Xi N)$, and $v_{cc'}$ are free-space interactions. The
starting channel $c_0$ correspond to $y=(\Lambda \Lambda )$. For
$y=(\Xi N)$ the Pauli operator $Q_{\Xi N}$ acts on intermediate
nucleon states. The energy denominator for $y\rightarrow y^{\prime
}$ transition is given by $e_{yy^{\prime }}$. For the intermediate
spectrum we adopt the so-called gap choice in which no potential
term is taken into account.

This coupled-channel treatment can be extended straightforwardly
to the $\Lambda \Lambda -\Xi N -\Sigma \Sigma$ three-channel case.
In the cases of using NHC-D, -F and NSC89, for simplicity, we
derive the $G$-matrix interactions in the $\Lambda \Lambda -\Xi N$
two-channel treatment. For the NSC97 models, however, we perform
the $\Lambda \Lambda -\Xi N -\Sigma \Sigma$ three-channel
calculations because of the following reason: The coupling
features of NSC97e and NSC97f are fairly different. The diagonal
potential derived from NSC97e in the two-channel treatment is
considerably more attractive than the corresponding one from
NSC97f, but they become similar to each other in the three-channel
treatment. Namely, the effect of $\Sigma \Sigma$ channel is (not)
substantial in the case of NSC97e (NSC97f). On the other hand, the
effects of $\Sigma \Sigma$ channels for the $\Lambda\Lambda-\Xi N$
effective coupling potentials are not so important both in the
cases of NSC97e and NSC97f.

The $G$-matrix interactions in coordinate space are represented by
Gaussian functions as follows: First we calculate the
momentum-space matrix elements  $\langle k|G_{\Lambda \Lambda
,\Lambda \Lambda }|k\rangle $ and $\langle k|G_{\Lambda
\Lambda,\Xi N }|k\rangle $ by solving Eq.\ (\ref{eq:Gmat}). Next,
we assume effective local potentials, which simulate the
calculated $G$-matrix elements, in three-range Gaussian forms:
\begin{equation}
\label{eq:eff}V(r)=a\exp (-r^2/r_1^2)+b\exp (-r^2/r_2^2)+c\exp (-r^2/r_3^2)
\end{equation}
with $r_1=1.5$ fm, $r_2=0.9$ fm, $r_3=0.5$ fm.
This Gaussian form is used both for the diagonal $\Lambda \Lambda$
interaction $V_{\Lambda \Lambda }$ and the $\Lambda \Lambda -\Xi N$
coupling interaction $V_{\Lambda \Lambda,\Xi N}$.
Parameters $a$, $b$ and $c$ are determined so that $\langle k|V%
_{\Lambda \Lambda}|k\rangle $ and $\langle k|V_{\Lambda \Lambda,\Xi N}
|k\rangle $ simulate the corresponding $G$-matrix elements $%
\langle k|G_{\Lambda \Lambda ,\Lambda \Lambda }|k\rangle $ and
$\langle k|G_{\Lambda \Lambda,\Xi N }|k\rangle $, respectively.
The $G$-matrices are calculated at a low density ($k_F=1.0$
fm$^{-1}$), because our concern in this work is light
double-$\Lambda $ hypernuclei. Here the uncertainty for choosing
the $k_F$ value is not so problematic in our analyses due to the
following reason:  The diagonal parts of our $G$-matrix
interactions are adjusted so as to reproduce $\Delta B_{\Lambda
\Lambda }(_{\Lambda \Lambda }^6\text{He})=1.0$ MeV, and the
$k_F$-dependences of their coupling parts are not so strong.

The old models NHC-D\ and NHC-F incorporate hard cores, which
radii $r_c$ can be treated as free parameters. We choose
$r_c=0.535$ fm (NHC-D) and $r_c=0.475$ fm (NHC-F) which give (for
the Isle-type $\Lambda-^{4}$He potential) $\Delta B_{\Lambda
\Lambda }(_{\Lambda \Lambda }^6\text{He})=0.90$ and 1.02 MeV,
respectively, rather close to $\Delta B_{\Lambda \Lambda
}(_{\Lambda \Lambda }^6\text{He})=1.0$ MeV \cite{Nag}. These hard
cores are entirely phenomenological and the results are quite
sensitive to their values. Here, some comment should be given
concerning the parametrization of the coordinate-space
interaction. With the hard-core models, the above procedure to
derive the three-range potential (\ref{eq:eff}) gives rise to the
strongly repulsive core in the $\Lambda \Lambda$ diagonal channel,
which is too singular to treat it in our HF model space. So, we
derived the version whose repulsive cores are comparable to those
for the cases of the soft-core models NSC89/97 with a sacrifice of
reproducing accurately the $k$-dependence of $\langle k|G_{\Lambda
\Lambda ,\Lambda \Lambda }|k\rangle $. The resulting core heights
are similar to those of the $\Lambda \Lambda$ $G$-matrix
interactions derived from NHC-D and -F according to another method
\cite{Yam94}.

On the other hand, the core parts of the Nijmegen soft-core models
are modeled more sophisticatedly on the basis of the SU(3)
symmetry. In the cases of NSC97 models, the potentials for $S=-2$
sector are determined with no additional free parameters
\cite{Stoks}. In the case of NSC89, it is necessary to choose the
$S$-wave form-factor mass in the $S=-2$ channel, though the
resulting potential is not so sensitive to this quantity: We take
1050.0 MeV rather tentatively. Then, the NSC89 model somewhat
overestimates $\Lambda\Lambda$ attraction ($\Delta B_{\Lambda
\Lambda }=1.43$ MeV in $_{\Lambda \Lambda }^6$He) whereas the
NSC97 models give a slight $\Lambda\Lambda$ repulsion: $\Delta
B_{\Lambda \Lambda }=-0.32$ MeV (e) and $\Delta B_{\Lambda \Lambda
}=-0.04$ MeV (f). In all the cases, we change the repulsive
short-range part $c$ in the diagonal $\Lambda \Lambda$ interaction
given by (\ref{eq:eff}) so as to reproduce the Nagara datum. Since
the existing models as well as data are far from being certain, we
do not consider the agreement/disagreement too seriously. Instead,
we compare the different potential models considering them as
examples of possible effective interactions motivated by
microscopic pictures, but phenomenological to some extent.

With the ORG-type $\Lambda$-core potentials, we adopt the same
procedure. Modifications required in the $\Lambda \Lambda $
diagonal potentials with respect to the Isle-type case are
typically small. Parameters of the diagonal $\Lambda \Lambda $ and
coupling $\Lambda \Lambda -\Xi N$ effective interactions are
presented in Table \ref{t2}.

\begin{table*}
\caption{\label{t2}Parameters of the diagonal
($V_{\Lambda\Lambda}$) and coupling ($V_{\Lambda\Lambda,\Xi N})$
potentials and volume integrals $\int V_{\Lambda \Lambda,\Xi
N}(r)\,d^3r$ for various potential models. All the entries are in
MeV except the rightmost column, which is in MeV\,fm$^3$.}
\begin{ruledtabular}
\begin{tabular}{lcccccccc}
Model&\multicolumn{4}{c}{$V_{\Lambda\Lambda}$}&
\multicolumn{4}{c}{$V_{\Lambda\Lambda,\Xi N}$}\\  & $a$ & $b$ &
$c$(Isle) & $c$(ORG)& $a$ & $b$&$c$&Vol.\ Int.
\\
\hline
NHC-D&$-5.659$&$-177.8$&925.0&916.0&0.1841&102.8&$-244.1$&250.9\\
NSC97f&$-5.380$&$-157.3$&810.0&808.0&1.361&109.2&$-193.5$&334.2\\
NSC97e&$-5.227$&$-168.7$&867.0&863.0&1.146&96.07&$-59.37$&370.2\\
NSC89&$-2.447$&$-98.60$&436.0&463.0&$-0.5035$&128.7&$-68.23$&465.5\\
NHC-F&$-1.768$&$-105.9$&462.0&488.0&$-0.9449$&199.3&$-300.5$&582.3\\
\end{tabular}
\end{ruledtabular}
\end{table*}

The last column in Table \ref{t2} shows the volume integrals for
the coupling potentials $\int V_{\Lambda \Lambda,\Xi N}(r)\,d^3r$.
This quantity reflects nicely the net strengths of the coupling
potentials. The strongest (weakest) coupling is seen to be
NHC-F(D). One should be careful in this case, however, that not
only the diagonal part, but also the coupling part depends on the
hard-core radius. As found in Table \ref{t2}, for instance, the
volume integral for NHC-F is 582.3 MeV\,fm$^3$. This strength is
obtained by taking $r_c=0.475$ fm, which leads to the reasonable
diagonal $\Lambda \Lambda$ attraction. When we take $r_c=0.53$ fm
(the same value as that in the $^1S_0$ $NN$ channel), the value of
the volume integral is reduced to 338.1 MeV\,fm$^3$.

Myint \textit{et al.}\ \cite{Ak03} used the simple Gaussian
potentials which are phase-shift equivalents to NSC97e, NHC-D and
NHC-F. In order to compare our method with theirs, let us derive
the $G$-matrix interactions from their phase-equivalent potential
to NSC97e.  The value of the volume integral of their coupling
potential in the isospin-singlet state is 525.2 MeV\,fm$^3$. The
corresponding value for the $G$-matrix version derived from their
potential is 401.8 MeV\,fm$^3$, which is not far from our value
370.2 MeV\,fm$^3$ for NSC97e in Table \ref{t2}. It is notable here
that the $G$-matrix coupling strength is demonstrated to be less
than the free-space one.

All $\Lambda \Lambda $ diagonal potentials used here are
fitted so as to reproduce the experimental value $%
\Delta B_{\Lambda \Lambda }(_{\Lambda \Lambda }^6\text{He})=1.0$
MeV \cite{Nag} using the single-channel approximation [equation
(\ref{e1}) without the coupling term]. From a general point of
view, it may be more reasonable to take into account the
$\Lambda\Lambda -\Xi N$ mixing explicitly not only in
$_{\Lambda\Lambda}^5$H ($_{\Lambda\Lambda}^5$He), but also in
$_{\Lambda\Lambda}^6$He. Previously, some estimations for the
$\Lambda\Lambda -\Xi N$ mixing effect in $_{\Lambda\Lambda}^6$He
were reported: For instance, Carr \textit{et al.}\ \cite{G97}
obtained the contribution to the binding energy about 0.2 MeV
unless $\Lambda\Lambda$ attraction is unreliably strong, while
Yamada and Nakamoto \cite{YaN} presented the contribution as large
as 0.4 MeV. Both the results were obtained using the NHC-D model
(or its simplified version) specified by a strong $\Lambda\Lambda$
attraction and a weak $\Lambda\Lambda -\Xi N$ coupling. The
difference between those results exhibits some uncertainties in
current approaches (generally, more elaborated than our one). It
can be deduced from \cite{G97,G03} that the coupling contribution
decreases if $\Lambda\Lambda$ attraction weakens, so probably
Nagara event \cite{Nag} implies a less value than previous
estimations. On the other hand, models with stronger
$\Lambda\Lambda -\Xi N$ coupling strengths may give more
substantial contributions.

Our approach is different from theirs on the basic point. Namely,
we renormalize contributions from high-lying $\Xi N$ states into
the diagonal $\Lambda\Lambda$ interaction through the $G$-matrix
procedure, and only low-lying $\Xi N$ states are treated
explicitly in our model space. In the cases of
$_{\Lambda\Lambda}^5$H and $_{\Lambda\Lambda}^5$He, there appears
the particular low-lying $\alpha+\Xi$ configuration. Then, the
intermediate nucleon is strongly bound in the $\alpha$ particle.
Such an intermediate state is forbidden by the Pauli principle for
a nucleon in the case of $_{\Lambda\Lambda}^6$He: Low-lying $\Xi
N$ states coupled to the $\Lambda\Lambda$ state are $\alpha+\Xi+N$
configurations, in which $N$ (and, probably, also $\Xi$) lies in
continuum. Energy differences of these intermediate $\Xi N$ states
from the $\Lambda\Lambda$ ground state are substantially greater
than that for the $\alpha+\Xi$ configuration. Similar
contributions from $^3\text{H(He)}+N+\Xi$ continuum configurations
occur also in $_{\Lambda\Lambda}^5$H ($_{\Lambda\Lambda}^5$He). In
principle, it would be reasonable to take into account the
continuum configurations in $_{\Lambda\Lambda}^6$He as well as in
$_{\Lambda\Lambda}^5$H ($_{\Lambda\Lambda}^5$He).

In our single-channel approximation for $_{\Lambda\Lambda}^6$He,
however, the ``diagonal'' $\Lambda\Lambda$ potential fitted to the
experimental binding energy incorporates effectively contributions
from $\Xi N$ intermediate states in which the nucleon is outside
from the $1s$ shell. The reasons to justify our procedure are as
follows: First, such contributions are expected to be small enough
due to large energy differences and small overlaps of wave
functions of the $\Lambda\Lambda$ bound state and $\Xi N$
continuum states. Secondly, these contributions in the cases of
$_{\Lambda\Lambda}^6$He and $_{\Lambda\Lambda}^5$H
($_{\Lambda\Lambda}^5$He) are supposed to be roughly equal to each
other and, therefore, to be simulated well by the diagonal
$\Lambda\Lambda$ interaction. Being simplified, our approach
enables us to avoid uncertainties arising from a treatment of the
coupling in $_{\Lambda\Lambda}^6$He.

If the coupling contribution to the binding energy of
$_{\Lambda\Lambda}^6$He is comparable to that from the diagonal
$\Lambda\Lambda$ potential, our results may be less reliable
quantitatively. However, there is no reason to expect that the
main effect (difference of the couplings in $_{\Lambda\Lambda}^5$H
and $_{\Lambda\Lambda}^5$He) can disappear even in this
unfavorable case. It should be emphasized that the coupling in
$_{\Lambda\Lambda}^6$He anyway deserves further careful study by
itself.

\section{\label{s3}Results and discussion}

First, we discuss the results obtained with Isle-type hyperon-nucleus
potentials.

In Table \ref{t3}, $\Delta B_{\Lambda \Lambda }(_{\Lambda \Lambda
}^5$H) and $\Delta B_{\Lambda \Lambda }(_{\Lambda \Lambda }^5$He)
calculated in the single-channel approximation without the
coupling are presented. Since all the diagonal $\Lambda \Lambda $
potentials are fitted to $\Delta B_{\Lambda \Lambda }(_{\Lambda
\Lambda }^6\text{He})=1.0$ MeV, they give also values close
to each other for $\Delta B_{\Lambda \Lambda }(_{\Lambda \Lambda }^5\text{H})%
=0.58$--0.63 MeV and
$\Delta B_{\Lambda \Lambda }(_{\Lambda \Lambda }^5\text{He})%
=0.65$--0.69 MeV. It is seen that even without the coupling,
$\Delta B_{\Lambda \Lambda }(_{\Lambda \Lambda }^5$He)$-\Delta
B_{\Lambda \Lambda }(_{\Lambda \Lambda }^5$H)$>$$0$. This nonzero
difference was first obtained in the five-body calculation
\cite{Nem} and then confirmed and explained in
\cite{FG2}. The origin of this difference is charge symmetry breaking $%
\Lambda N$ interaction. Since $B_\Lambda (_\Lambda
^4$He)$-B_\Lambda (_\Lambda ^4$H)$>$$0$, $\Lambda $ hyperons in
$_{\Lambda \Lambda }^5$He move closer to the center and,
therefore, closer to each other than in $_{\Lambda \Lambda }^5$H.
So they attract each other somewhat stronger in $_{\Lambda \Lambda
}^5$He than in $_{\Lambda \Lambda }^5$H \cite{FG2}. This
difference is less than 0.1 MeV (whereas the difference in the
$B_\Lambda$ values in the corresponding single-$\Lambda$
hypernuclei is about 0.3 MeV) if the $\Lambda \Lambda $ attraction
is compatible with Nagara event and can be greater for stronger
$\Lambda \Lambda $ attraction, but not greater than several tenths
of MeV \cite{FG2}. It is seen that the coupling effect increases
the difference considerably (columns labeled cc in Table \ref{t3},
corresponding to the Xa1 potential.).

\begin{table}
\caption{\label{t3}Single-channel (sc) and coupled-channel (cc)
$\Delta B_{\Lambda\Lambda}$ values in MeV calculated from eq.\
(\protect\ref{e3}) with $B_\Lambda(_\Lambda ^4\text{H})=1.25$ MeV
and $B_\Lambda(_\Lambda ^4\text{He})=1.53$ MeV  for various
potential models. In the coupled-channel calculation, the Xa1
potential is used.}
\begin{ruledtabular}
\begin{tabular}{lcccc}
Model&\multicolumn{2}{c}{$\Delta
B_{\Lambda\Lambda}(^5_{\Lambda\Lambda}$H)}
&\multicolumn{2}{c}{$\Delta
B_{\Lambda\Lambda}(^5_{\Lambda\Lambda}$He)}\\  &sc&cc&sc&cc\\
\hline NHC-D&0.63&0.72&0.69&0.84\\ NSC97f&0.62&0.79&0.68&0.95\\
NSC97e&0.63&0.85&0.69&1.05\\ NSC89&0.58&0.97&0.65&1.25\\
NHC-F&0.58&1.16&0.65&1.55\\
\end{tabular}
\end{ruledtabular}
\end{table}

\begin{figure*}
\includegraphics*[width=75mm]{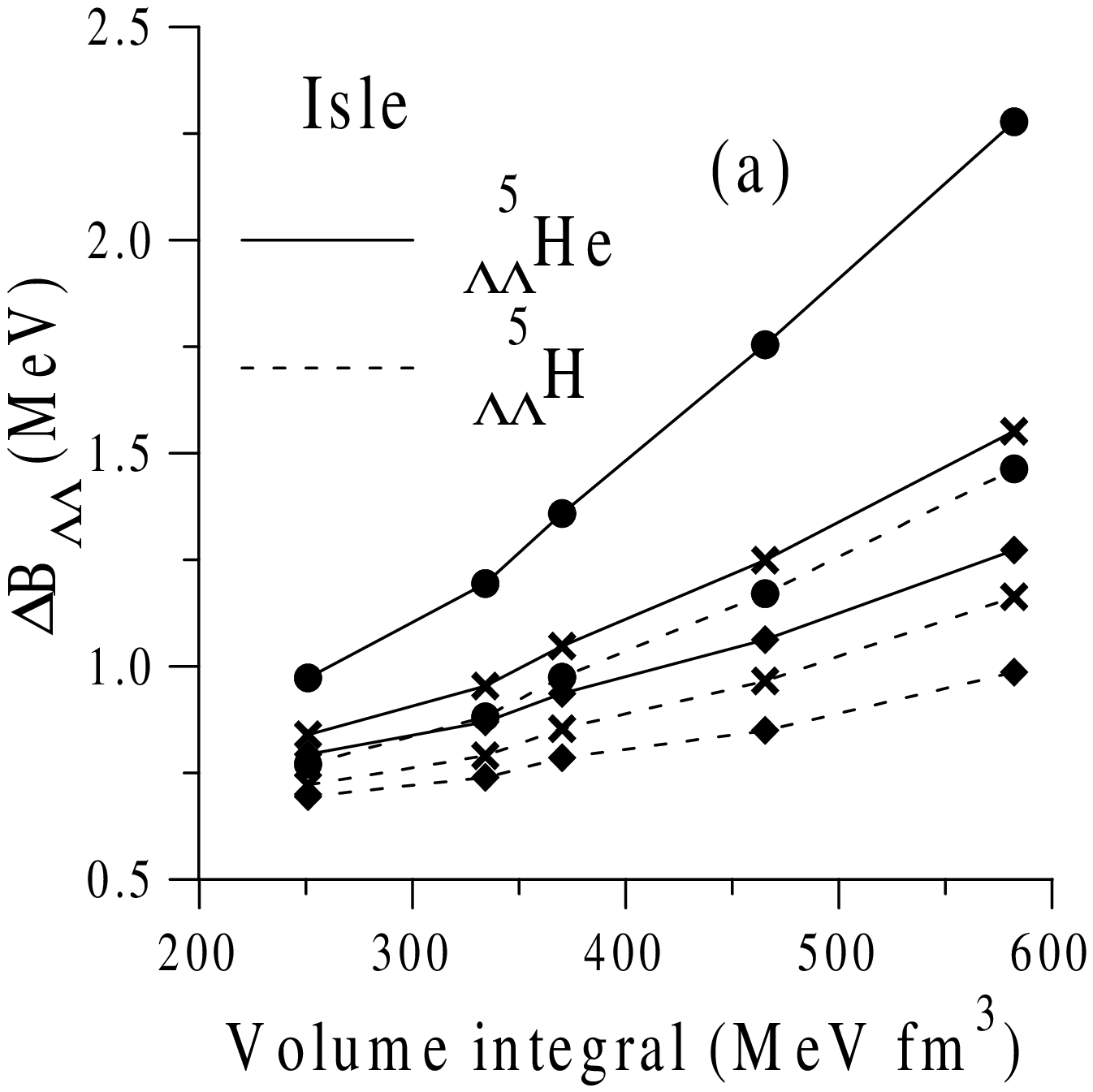}
\includegraphics*[width=75mm]{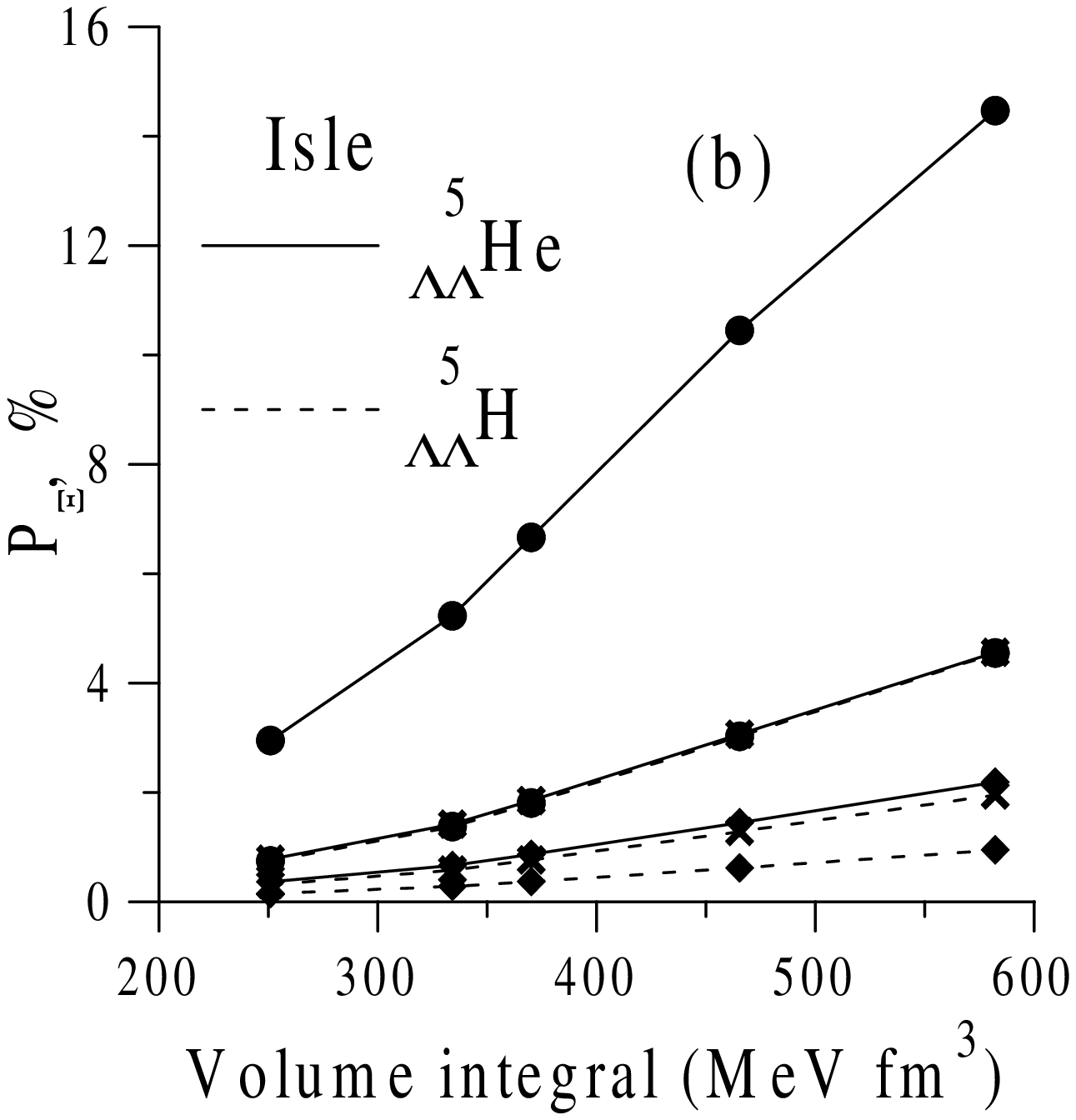}
\caption{\label{f1}$\Delta B_{\Lambda\Lambda}$ (a) and $\Xi$
admixture probabilities $p_\Xi$ (b) as functions of volume
integral $\int V_{\Lambda\Lambda,\Xi N}\,d^3r$ in $_{\Lambda
\Lambda }^5$H and $_{\Lambda \Lambda }^5$He for $\Xi\alpha$
Isle-type potentials Xa1 (crosses), Xa2 (circles), and Xa0
(diamonds) and the Isle-type $\Lambda$-core potentials. Solid
($_{\Lambda \Lambda }^5$He) and dashed ($_{\Lambda \Lambda }^5$H)
lines are drawn as a guide for eyes. Diagonal $\Lambda\Lambda$ and
coupling $\Lambda\Lambda-\Xi N$ potentials are (from left to
right) NHC-D, NSC97f, NSC97e, NSC89, and NHC-F.}
\end{figure*}

In Fig.\ \ref{f1}(a), $\Delta B_{\Lambda \Lambda }$ values
obtained from the full calculation with various Isle-type
$\Xi\alpha$ potentials are shown as functions of volume integral
$\int V_{\Lambda\Lambda,\Xi N}(r)\,d^3r$.

It is seen that the coupling effect is anyway meaningful and may
be rather high. Even with moderate $\Xi \alpha $ potential Xa1,
full $\Delta B_{\Lambda \Lambda }$ is more than twice as large as
the single-channel
value for strong coupling interactions. For the NHC-F model, the difference $%
\Delta B_{\Lambda \Lambda }(_{\Lambda \Lambda }^5$He)$-\Delta B_{\Lambda
\Lambda }(_{\Lambda \Lambda }^5$H) is about 0.4 MeV. For the strongest $\Xi
\alpha $ potential Xa2, $\Delta B_{\Lambda \Lambda }(_{\Lambda \Lambda }^5$He)
can reach 2.3 MeV (remind that $\Delta B_{\Lambda \Lambda }(_{\Lambda
\Lambda }^6$He)$=1.0$ MeV). Even for zero $\Xi \alpha $ potential Xa0, $%
\Delta B_{\Lambda \Lambda }(_{\Lambda \Lambda }^5$He) can exceed
$\Delta B_{\Lambda \Lambda }(_{\Lambda \Lambda }^6$He) as has been
pointed out in \cite{Ak03}.

The $\Xi $ admixture probability $p_\Xi =\int_0^\infty \chi ^2\,
dr$ is presented in Fig.\ \ref{f1}(b). Its dependence on the
volume integral follows closely the corresponding curves for
$\Delta B_{\Lambda \Lambda }$. Most of the values in Fig.\
\ref{f1}(b) exceeds considerably $p_\Xi $ obtained in $\Lambda
\Lambda $ hypernuclei with meson-exchange models earlier. In the
extreme case of the strong coupling and strong $\Xi \alpha $
potential, $p_\Xi $ reaches 14\%. The same value has been obtained
in $_{\Lambda \Lambda }^{10}$Be \cite{YaN} in a quark model with
strong dibaryonic mixing.

It is seen that $\Delta B_{\Lambda \Lambda }$ and $p_\Xi $ are
smooth functions of the volume integral. At relatively weak
couplings, $\Delta B_{\Lambda \Lambda }$ and $p_\Xi $ are nearly
quadratic in the volume integral according to the lowest order of
perturbation theory. Generally the coupling effect depends on the
coupling potential strength (volume integral) rather stronger than
on its shape.

The coupling contributions to $\Delta B_{\Lambda \Lambda }$'s in
our calculations with no $\Xi\alpha$ potential are systematically
smaller than those obtained by Myint \textit{et al.}\ \cite{Ak03}.
The numerical comparison between the two approaches can be
exemplified for the NSC97e model. The $G$-matrix interaction
derived from their simplified version (97e) is similar
qualitatively to that from the original NSC97e, as mentioned
before. The obtained values of $\Delta B_{\Lambda \Lambda}$ for
$^{\ 5}_{\Lambda \Lambda}$H and $^{\ 5}_{\Lambda \Lambda}$He are
0.96 MeV and 1.28 MeV, respectively \cite{Ak03}. These values are
considerably larger than our obtained values not only with Xa0
(0.79 and 0.94 MeV), but also with Xa1 (see Table \ref{t3}).

Certainly, our approach differs from theirs by many features.
Here, it is worthwhile to comment critically the method applied by
Myint \textit{et al.}\ to reduce the two-channel problem to the
single-channel one. For this aim, the effective $\Lambda\Lambda$
diagonal potential
\begin{equation}
V_{\Lambda \Lambda }^{\text{eff}}=V_{\Lambda \Lambda }-V_{\Lambda
\Lambda,\Xi N} \frac 1{\overline{\Delta E}}V_{\Xi N,\Lambda
\Lambda}\label{e8}
\end{equation}
is used incorporating the coupling due to the second term (for
brevity, we do not write down separately the terms corresponding
to the $\Xi^-p$ and $\Xi^0n$ channels). Essential approximation of
Myint \textit{et al.}\ is replacement of the full energetical
denominator by average constant $\overline{\Delta E}$, which is
adjusted to $\Lambda\Lambda$ scattering properties calculated in a
meson-exchange model.

The contribution of the second term of (\ref{e8}) to a
$\Lambda\Lambda$ hypernuclear energy can be easily expressed as
\begin{equation}
\epsilon =-\frac{\sum_i\left| \left\langle \phi \left| V_{\Lambda
\Lambda, \Xi N}\right| \chi _i\right\rangle \right|
^2}{\overline{\Delta E}},\label{e9}
\end{equation}
where $\phi$ is the wave function of the first ($\Lambda\Lambda$)
channel solved with $V_{\Lambda \Lambda }^{\text{eff}}$, and
$\left|\chi_i\right>$ represents the complete set of states of the
second ($\Xi N$) channel.

On the other hand, the accurate expression for the second-order perturbative
contribution, considering the Pauli suppression effect on nucleon occupied
states, is given as
\begin{subequations}
\label{e101}
\begin{eqnarray}
\epsilon &=& -\sum_{i \notin P}\frac{\left| \left\langle \phi
_0\left| V_{\Lambda \Lambda,\Xi N}\right| \chi _i\right\rangle
\right| ^2}{E_i-E_0} \label{e10}  \\*
&=& -\sum_{i}\frac{\left|
\left\langle \phi _0\left| V_{\Lambda \Lambda,\Xi N}\right| \chi
_i\right\rangle \right| ^2}{E_i-E_0}
 +\sum_{i \in P}\frac{\left| \left\langle \phi _0\left|
V_{\Lambda \Lambda,\Xi N}\right| \chi _i\right\rangle \right|
^2}{E_i-E_0}\label{e11}
\end{eqnarray}
\end{subequations}
where $\phi_0$ and $E_0$ correspond to the unperturbed (uncoupled)
state and $P$ denotes Pauli forbidden $\Xi N$ states. The
summation in the first term in (\ref{e11}) runs over all the $\Xi
N$ states and corresponds to (\ref{e9}). The second term comes
from the Pauli correction, corresponding to $\Delta
V_{\text{Pauli}}$ in \cite{Ak03}. Since the $\Xi$ admixtures are
typically not so high, the lowest order of perturbation is
appropriate at least qualitatively; the difference between $\phi$
and $\phi_0$ is probably small. To reduce the first term of
(\ref{e11}) to (\ref{e9}), one should assume some averaged value
for the denominator. Clearly, this averaged value is defined by
the $\Xi$ hypernuclear spectrum and bears no relation to the
two-baryon c.m.\ averaged energy $\overline{\Delta E}$ adjusted so
as to simulate the $\Lambda\Lambda$ scattering parameters. On the
other hand, Myint \textit{et al.}\ \cite{Ak03} use the energy
denominators conforming to (\ref{e101}) in the calculation of
many-body corrections ($\Delta V_{\text{Pauli}}$ and $\Delta
V_{\text{alpha}}$ in their notations). Conceptually, in
(\ref{e11}) $\overline{\Delta E}$ is used in the first term, but
not in the second term. It leads to the contradiction pointed out
by Filikhin \textit{et al.}\ \cite{FGS}: The correction introduced
to exclude the Pauli forbidden state, if calculated properly, is
greater than the whole coupling effect. This contradiction is
considered to be originated from the inconsistent treatments for
the two terms in Eq.\ (\ref{e11}), which are obtained only by
reforming the single term (\ref{e10}). Namely, their treatment
does not accord to the clear-cut expression (\ref{e101}) and the
obtained results are considered to be questionable.

We stress also the role of the $\Xi \alpha $ potential omitted in
\cite{Ak03}. Naturally, the stronger is $\Xi \alpha $ attraction,
the greater is the coupling. Substitution of zero $\Xi \alpha $
potential Xa0 by even relatively weak
attractive Xa1 gives an energy gain up to 0.2 MeV in $_{\Lambda \Lambda }^5$%
H and 0.3 MeV in $_{\Lambda \Lambda }^5$He. For the Xa2 potential,
the gain can reach 0.5 MeV ($_{\Lambda \Lambda }^5$H) and 1.1 MeV
($_{\Lambda \Lambda }^5$He). The $\Xi \alpha $ potential is found
to amplify the effect, making it observable more simply. On the
other hand, it is clear that reliable quantitative extraction of
the coupling strength from future data is improbable until one
deduces reliable $\Xi \alpha $ potential.

Qualitatively, the effect of the $\Xi \alpha $ potential can be described as
follows. Introducing $\Xi \alpha $ attraction, one essentially moves the
unmixed $_\Xi ^5$Z state closer to the unmixed $_{\Lambda \Lambda }^5$Z
state. Naturally, the smaller is the energy difference, the greater is the
mixing. Seemingly, similar effect can be provided by changing of the $%
\Lambda \Lambda $ diagonal potential. But for the $\Lambda \Lambda
$ potential, this is not the case. If the diagonal $\Lambda
\Lambda $ interaction increases (the $_{\Lambda \Lambda }^5 $Z
level moves down and away from the $_\Xi ^5$Z one), the coupling
can even grow. The reason is that the weaker is $\Lambda \Lambda $
attraction, the more extended is $\Lambda $ spatial distribution.
Therefore, overlap of the $\Lambda $ wave function with the
nucleonic one becomes poorer (note that the nucleon is bound in
the $\alpha $ particle rather strongly) and the coupling strength
is reduced. This factor can overcome the decrease of the initial
energy difference. Some schematic numerical illustrations have
been presented in \cite{L03}.

\begin{figure*}
\includegraphics*[width=75mm]{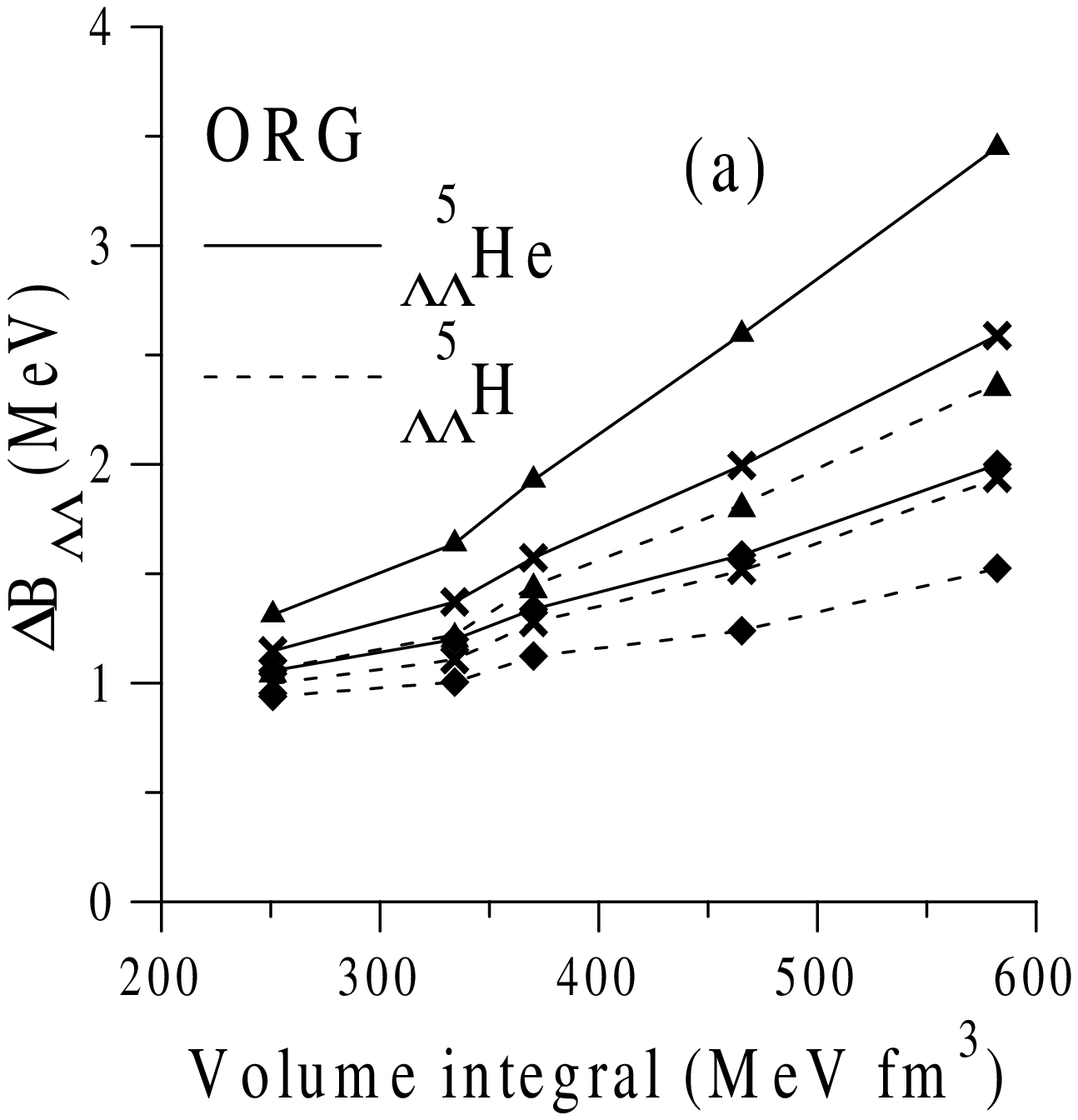}
\includegraphics*[width=75mm]{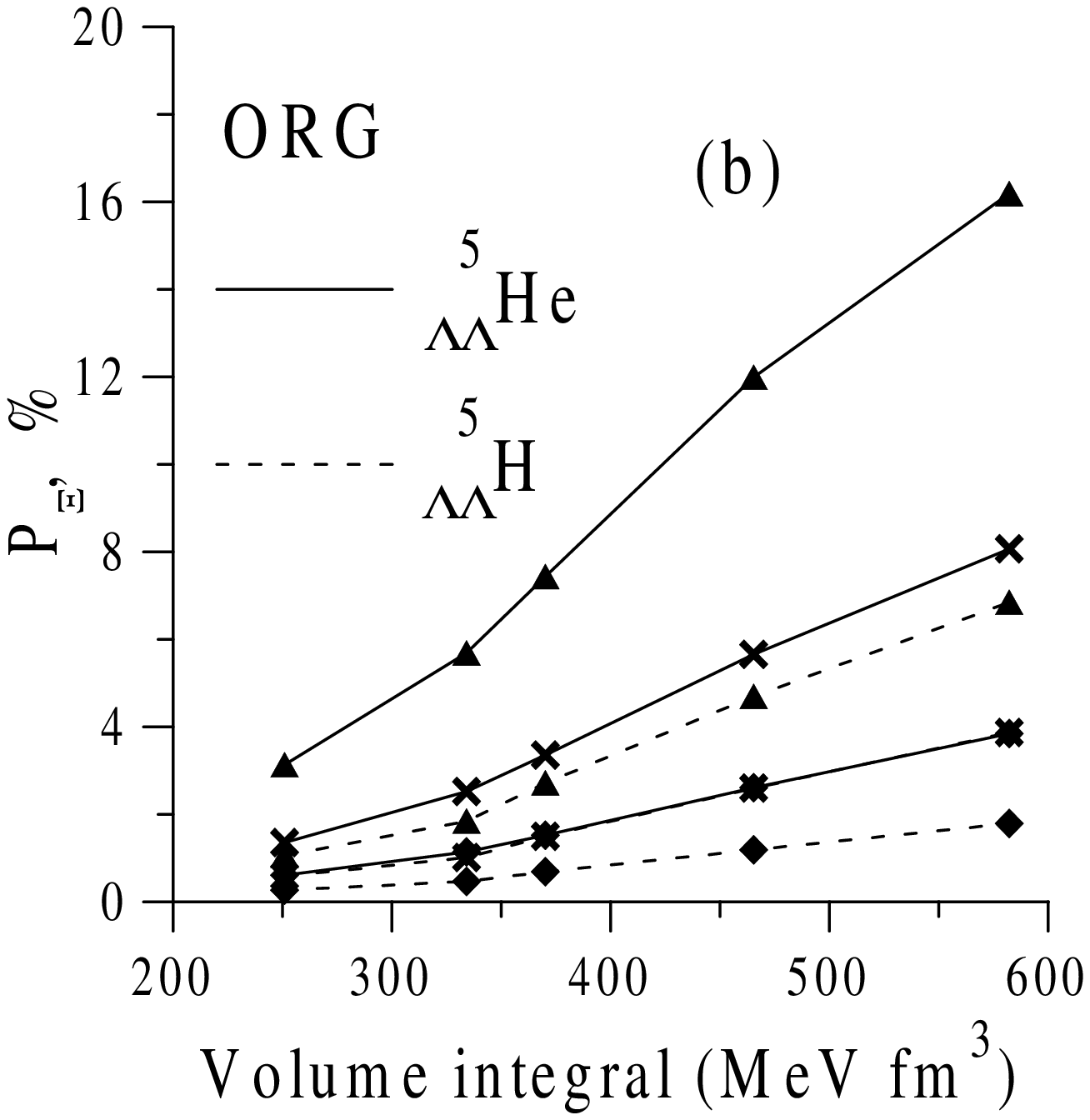}
\caption{\label{f2}The same as in Fig.\ \protect\ref{f1} for the
ORG-type $\Xi\alpha$ potentials Xa3 (crosses), Xa4 (triangles),
and Xa0 (diamonds) and the ORG-type $\Lambda$-core potentials.}
\end{figure*}

Here we demonstrate another example of the importance of the
hyperonic
spatial distributions. We repeat the calculations using ORG-type $%
U_{c\Lambda }$ and $U_\Xi $. The $\Delta B_{\Lambda \Lambda }$
values are shown in Fig.\ \ref{f2}(a) and $\Xi $ probabilities are
presented in Fig.\ \ref{f2}(b). It is seen that the coupling
effect increases considerably with respect to the Isle-type
potential form. In the extreme cases, $\Delta B_{\Lambda \Lambda
}(_{\Lambda \Lambda }^5$He) reaches huge values, almost five times
as $\Delta B_{\Lambda \Lambda }(_{\Lambda \Lambda }^5$He) from the
single-channel calculation (remind that ``extreme'' ORG-type $\Xi
\alpha $ potential Xa4 is weaker than Xa2 of the Isle-type). The
$\Xi $ admixture probabilities are larger too, though to a less
extent.

\begin{figure}
\includegraphics*[width=85mm]{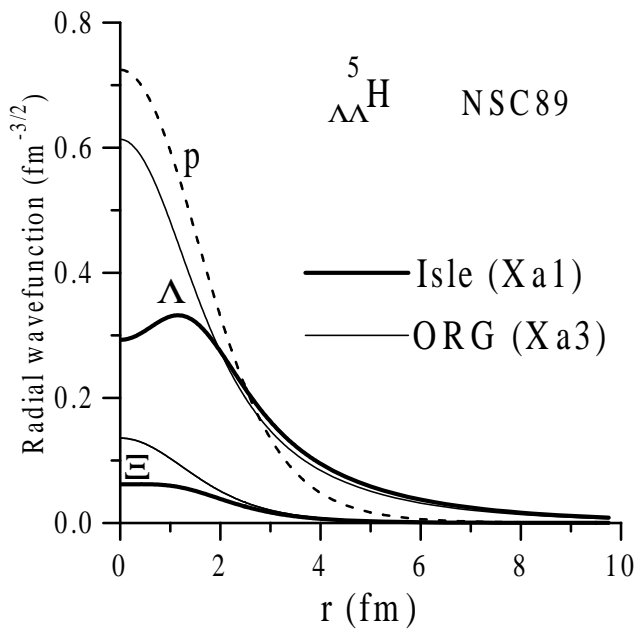}
\caption{\label{f3}Radial wave functions $\varphi(r)/r$ and
$\chi(r)/r$ for $_{\Lambda \Lambda }^5$H with the NSC89 potential
and the Xa1 and Xa3 potentials. Thick lines correspond to the
Isle-type potentials, thin ones are for the ORG-type potentials.
The dashed line shows proton wave function $\varphi_N(r)/r$.}
\end{figure}

The reason for this enhancement is similar to that discussed
above. The ORG-type $\Lambda$-core potential gives much more
concentrated hyperon spatial distributions than the Isle-type
potential does. It is seen from Fig.\ \ref{f3}, where the
hyperonic radial wave functions in $_{\Lambda \Lambda }^5$H are
shown in comparison with the proton wave function. Therefore, the
overlap becomes larger, and the mixing matrix element $\sigma $ in
(\ref{e1}) and (\ref{e2}) increases. It is interesting that
replacement of the strongest ORG-type Xa4 potential by still
stronger Isle-type potential Xa2 does not enhance further the
coupling since the shapes of the $\Lambda -{}^3$Z and $\Xi \alpha
$ potentials become inconsistent.

Usually, the Isle-type potentials are treated as more realistic for light
hypernuclei. However, it is instructive that there exists such a high
sensitivity of the coupling effect to the shape of the $\Lambda $-core
and $\Xi \alpha $ potentials. Though the coupling effect in the calculation
with
the ORG-type potentials is probably overestimated, it is still possible that
the true effect is somewhat larger than that obtained with the Isle-type
potentials.

\section{\label{s4}Conclusion}

We studied properties of mirror hypernuclei $_{\Lambda \Lambda }^5$H and $%
_{\Lambda \Lambda }^5$He in the two-channel approach with $G$-matrix
effective interactions. The $\Lambda \Lambda -\Xi N$ coupling is
particularly efficient in these hypernuclei owing to
the small energy-differences between the $\Lambda \Lambda$ and
$\Xi N$ states. Most importantly, the coupling in $_{\Lambda \Lambda }^5$He
is substantially larger than in $_{\Lambda \Lambda }^5$H.
The $\Lambda \Lambda -\Xi N$ coupling leads to considerable
difference of the binding energies of $_{\Lambda \Lambda }^5$He and $%
_{\Lambda \Lambda }^5$H. This difference hardly can be explained by any
other reasons and can be a clear signature of the baryonic mixing. Possibly,
it is the most unambiguous signature of baryonic mixing among various
suggestions considered so far for ordinary nuclei as well as hypernuclei.

Essentially, the effect analyzed here is a charge symmetry
breaking effect of an unusual (many-body) nature. Namely, the
$\Lambda \Lambda $ interaction in the mirror $\Lambda \Lambda $
hypernuclei appears to be different due to electromagnetic mass
difference of $\Xi ^{-}$ and $\Xi ^0$, since different nucleonic
states in the mirror cores are occupied. Another charge symmetry
breaking mechanism for free $\Lambda N$ interaction, originating
from the mass difference of $\Sigma ^{+}$ and $\Sigma ^{-}$, has
been suggested by Dalitz and von Hippel \cite{DvH} (for further
studies, see \cite{GGW,NKG}). By contrast, our mechanism is
essentially many-body, existing only due to the nuclear
environment.

Our analysis demonstrates that detailed knowledge of the $\Xi
\alpha $ and $\Lambda -{}^3$Z potentials is needed in order to
extract the $\Lambda \Lambda -\Xi N$ coupling strength
quantitatively from future data on $_{\Lambda \Lambda }^5$H and
$_{\Lambda \Lambda }^5$He. Whereas the $\Lambda$-nucleus
interaction, though being far from complete understanding, is
recognized to a large extent, $\Xi $-nucleus interaction is known
quite poorly at present. Our consideration exhibits that generally
there are no separate branches of $\Lambda \Lambda $ and $\Xi $
hypernuclear dynamics, but rather the unified field of $S=-2$
hypernuclei. We showed the example when not only
the $\Lambda \Lambda -\Xi N$ coupling interaction, but also the $\Xi $%
-nucleus diagonal potential, is essential for $\Lambda \Lambda $
hypernuclear properties. On the other hand, the same $\Lambda \Lambda -\Xi N$
potential is responsible for conversion widths of $\Xi $ hypernuclei whereas
branching ratios of conversion channels depend substantially on the $\Lambda
\Lambda $ potential \cite{Yam97}.

Due to the small mass difference of the $\Xi N$ and $\Lambda
\Lambda $ pairs, the $\Xi N$ admixtures in $\Lambda \Lambda $
hypernuclei are not only important, but rather its importance is
drastically different in different hypernuclei (and, probably, in
different states). Evidently, the $\Xi N$ mixing in the
five-baryon $\Lambda \Lambda $ hypernuclei is considerably larger than in $%
_{\Lambda \Lambda }^6$He. It was suggested that the
$\Lambda \Lambda -\Xi N$ coupling effect in $_{\Lambda \Lambda }^4$H is
also large \cite{G94}. It should be noted, however, that
baryonic spatial distributions in $_{\Lambda \Lambda }^4$H are expectedly
rather extended. We showed here that it is an unfavorable factor for the
coupling. This problem is very interesting in view of the question whether $%
_{\Lambda \Lambda }^4$H is bound or not, which was answered oppositely in
two recent studies \cite{FG3,Ak02} basing on comprehensive four-body
(however, single-channel) calculations.

Lastly, some comments on the possible observation of the five-body $\Lambda
\Lambda $ hypernuclei are in order. The $_{\Lambda \Lambda }^5$H
hypernucleus can be produced both from $\Xi ^{-}$ capture reactions \cite
{Yam97,Yam942} or the $^7$Li$(K^{-},K^{+})$ reaction \cite{Ak96} and
detected by characteristic $\pi ^{-}$ emission \cite{Yam97} though detection
by the $\Lambda \Lambda \rightarrow \Sigma ^{-}p$ weak decay suggested in
\cite{Ak96} is problematic in view of recent calculations \cite{Ito,Par,Jun}
predicting very small branching ratios for the $\Sigma $ emission channel.
On the other hand, production rates of $_{\Lambda \Lambda }^5$He from $\Xi
^{-}$ capture by various nuclei are much smaller \cite{Yam97,Yam942}.
Possibly, $_{\Lambda \Lambda }^5$He can be produced with a detectable
probability from $\Xi ^{-}$ capture by lithium. However, the detection of $%
_{\Lambda \Lambda }^5$He is rather difficult since its $\pi ^{-}$ decay is
expectedly rather improbable. The technique to detect $_{\Lambda \Lambda }^5$%
He is, therefore, an open problem.

\begin{acknowledgments}
The work of one of the authors (D.\ L.) is supported in part by
Russian National Program for support of leading scientific schools
(grant No. 1619.2003.2).
\end{acknowledgments}

\end{document}